\documentclass{article}

\PassOptionsToPackage{numbers, compress}{natbib}

\usepackage[final]{neurips_2023}

\usepackage[utf8]{inputenc} 
\usepackage[T1]{fontenc}    
\usepackage{hyperref}       
\usepackage{url}            
\usepackage{booktabs}       
\usepackage{amsfonts}       
\usepackage{nicefrac}       
\usepackage{microtype}      
\usepackage{xcolor}         
\usepackage{graphicx}
\usepackage{cleveref}
\usepackage{appendix}
\usepackage{color, colortbl}

\title{Foundational Moral Values for AI Alignment}

\author{%
  Betty Li Hou \\
  New York University \\
  \texttt{blh9134@nyu.edu} \\
  \And
  Brian Patrick Green \\
  Markkula Center for Applied Ethics \\
  \texttt{bpgreen@scu.edu} \\
}

\begin{document}

\maketitle

\begin{abstract}

  Solving the AI alignment problem requires having clear, defensible values towards which AI systems can align. Currently, targets for alignment remain underspecified and do not seem to be built from a philosophically robust structure. We begin the discussion of this problem by presenting five core, foundational values, drawn from moral philosophy and built on the requisites for human existence: survival, sustainable intergenerational existence, society, education, and truth. We show that these values not only provide a clearer direction for technical alignment work, but also serve as a framework to highlight threats and opportunities from AI systems to both obtain and sustain these values. 
  
\end{abstract}

\section{Introduction}

Humanity and human societies are inherently technological. Our dependence on technology is a human universal: found in all human societies \citep{brown1991human} and as a standard part of human psychology starting in early childhood \citep{casler2005young}. Technology and society typically work in concert in order to create a functioning technological culture, but as technology advances faster than culture can adapt, society can be put at risk of harm due to the misalignment of technological means to social ends. This issue with respect to AI is the AI alignment problem, and it is one of the most important ethical questions in our contemporary world. It confronts us with the questions: In general, what is the proper relationship between humanity and technology? And, specifically, how do we live in a fully sustainable society with new, increasingly capable AI? 

Alignment can be defined as ensuring that an AI system does what it is intended to do; this might include avoiding harm and suffering to individuals and mitigating catastrophic risks to society \citep{nick2014superintelligence, leike2018scalable, russell2019human}. Underlying these is the question of what target AI should align to--what and whose intentions do we align to \citep{kirk2023signifier}, and what does avoiding harm, suffering, and catastrophe entail? One way to look at this is through moral values. If we can define a set of fundamental moral values, then these values can help guide the development of AI models towards social benefit. 

Additionally, AI alignment is not a problem which occurs on only one level: it is a multilevel problem occurring at the individual, organizational, national, and international levels \citep{hou2023multi}. Any solution to the alignment problem must take into account its multilevel nature; solutions which do not include all levels cannot be complete solutions. Humans do not live as isolated individuals, but as groups, nations, and the entire world. AI alignment must include the product level, such as LLMs \citep {kenton2021alignment} (within the organizational level, i.e. corporations), and it must also be solved at every level. Therefore, we need to think about alignment as a global problem of the governance, development, deployment, and usage of AI rather than solely trying to align a system alone \citep{hou2023multi, choung2023multilevel}. These multiple levels are the \emph{form} of the solution, but the solution also requires \emph{content}. If AI is to be aligned, it must be aligned with something, i.e., moral values. 

Iason Gabriel describes the alignment problem as a combination of a technical and normative problem, the technical looking at “how to formally encode values or principles in artificial agents so that they reliably do what they ought to do", and the normative asking what values or principles, if any, we ought to encode in artificial agents \citep{gabriel2020artificial}. The technical side of the problem is intrinsically linked with the normative side as one cannot be solved without the other. Without goals from the normative side, we would have a means for alignment, but no end. Furthermore, in order of prioritization, survival must precede flourishing. In other words, we can’t think about the questions of “What values do we encode into AI for human flourishing?” without having satisfactorily answered “What do we need to consider about AI in order for humans to survive?” The latter can then inform us when looking at the former. Following this logic, at some point the values that AI aligns to must hit bedrock foundational values. These are core values of human existence, related to our nature as a species, and are the basis for more specified moral values and principles.

Hendrycks et al., Awad et al., Tasioulas, Sorensen et al., and others have previously explored issues related to aligning AI with values, and what we propose here, being at the very foundational level, is not incompatible with these previous approaches and in fact can provide them additional support \citep{hendrycks2020aligning, awad2022computational, tasioulas2022artificial, sorensen2023value}. In this paper, we propose five foundational values with which to align AI and lay out an agenda that follows from the values. These values are naturalistic \citep{flanagan2016naturalizing} and logically grounded in human survival as a precondition for ethics \citep{jonas1984imperative}. As we describe them here, these five values are not only the bare minimum for human survival, but also, if construed more maximally, the foundation for human individual and social flourishing. Solving the alignment problem should not only seek to protect human sustainable survival, but also seek to facilitate a good future for humankind, with AI broadly beneficial to all.

\section{Five Foundational Moral Values}

Coming up with a set of universally agreed-upon values is a daunting but necessary undertaking, and human rights discourse \citep{shue2020basic, donnelly2013universal}, capabilities theory \citep{nussbaum2000women}, etc., have already done much to help in this regard. However, with respect to AI, in many cases attempts to answer the question of “To what values do we align?” have been left underspecified. Recent alignment work in LLMs has emphasized three values: helpfulness, harmlessness, and honesty \citep{askell2021general, bai2022training}--however, what each of these mean under various conditions remains hard to say. We are immediately confronted by the question: “helpfulness” and “harmlessness” with respect to what? And what exactly is meant by honesty? As these questions do not have clear answers yet suggests that the values alone may be insufficient for both the most abstractly foundational and the most concrete practical and critical situations. Building a fuller ethical structure--with foundational moral values, mid-level principles, and very concrete action-guiding principles--can help in these cases \citep{flahauxethics}. These deeper, foundational values are the ones we explore here and aim to bring attention to. 

The five fundamental moral values we present here are irreducible in that they exist as a network. Each depends on the others; even a principle as fundamental as survival relies upon truth, for example, and vice versa. These fundamental values are also necessary: without them human existence would be impossible, not to mention AI alignment with humanity. These values are social in that they apply to groups of people rather than individuals, though certainly individuals are the necessary basis upon which groups exist. Without them a human group will not survive. Philosophical defenses of these values will be given below. As this task is again a big undertaking, future work should suggest alternative foundational values for comparison. 

In order for humanity to exist, there are several foundational conditions which constitute moral values that humans need to seek.

\begin{enumerate}
    \item Humans need to \textbf{survive}. As the foundation for the existence of AI and alignment at all, human survival is necessary. This entails looking for food, having shelter, maintaining one’s health, warding off predators, preparing for or preventing disasters, and so on. 
    \item Humans need to survive long term. As a species that is born, ages, and dies, mere short-term survival is not enough, we also need to reproduce so that future generations can continue to exist, and protect our environment so that future generations can live. We refer to this as \textbf{sustainable intergenerational existence}. 
    \item Humans need to live in \textbf{society}. We cannot survive as individuals: we need families to raise us as children, we need others to care for us when we are sick, we need goods from other locales, and so on. Society is a structure for individuals to coexist in. In our present day, this means division of labor, the entire economic supply chain, resource management, and all the knowledge and skills necessary to live in society. 
    \item Humans need to live in society long term. Following our nature as generational beings, eventually expert humans die and need to be replaced by younger, educated humans. In other words, humans need \textbf{education}, both in the formal sense of schooling as well as informally learning about culture and society. Because humans lack the detailed instincts of other living creatures, we need culture to replace instinct, and the conveyance of culture between generations requires education and training of the young in the ways of their culture. 
    \item Humans need to know true facts (\textbf{truth}) about reality, or at least our understanding of reality needs to be “true enough” in order to be certain enough of what we know and what we teach. While for much of human history our ancestors might not have known very much, they knew enough to stay alive, live sustainably across generations, live in society, educate their children, and continue the existence of the human species (which might be more than we can say for our present time). Truths are both theoretical and practical, and while people of the past might have been short on theoretical truths, they had sufficient practical ones to give us our chance, in this generation.
\end{enumerate}

Not all individuals must do all five conditions, as these values are presented for groups of humans. People can die in war to protect others, not everyone has to reproduce, not everyone must be integrated in a society all the time, etc., but collectively we must do all five if we are to continue to exist as a species. When looking at AI alignment however, a powerful AI system should be aligned to all five conditions as its scope of impact is potentially the entire human species, even if not directly but rather through the whole range of downstream impacts.

These values can also be characterized positively or negatively, e.g., seek survival or avoid extinction; live in society or avoid living in isolation; seek truth or avoid falsehood. This framing can be important when thinking about particular issues with alignment, as they can change the scope significantly.

Importantly, these values are demonstrable by \textit{reductio ad absurdum}. Choose to value the opposite of any one of the values and humans go extinct, thus undercutting the precondition for valuation at all: humans to do the valuing. E.g., to value and seek extinction would negate the possibility of moral valuations at all, thus absurdly valuing not valuing. For additional description see Appendix \ref{appendix:phil}.

\section{The Five Foundational Moral Values and Philosophy}

The above helps show that these five foundations are necessary values for humankind to exist (even if implicit, or practiced, but not explicitly understood in theory \citep{rappaport1979ecology}), but it is also worth noting that many philosophers from around the world and various times have also advocated for these five foundations in their cultural contexts. Here we will provide a very quick overview, concentrating on Aristotle, Confucius, and a few other more contemporary thinkers or practical behaviors.

\paragraph{Survival} The survival of humanity seemed like a given for most of history and so this value is typically implicit and underlies the assumptions of various philosophers. For example, Aristotle assumed that humanity had lived infinitely into the past and thus would also live indefinitely into the future. This was an assumption, based on his eternal metaphysics, and therefore immediate extinction risks were unthinkable \citep{aristotle1984complete, moynihan2020x}. Fei Teng explains that Confucius can also be used to argue for a form of intergenerational justice based on virtue ethics as a response to climate change, and this can be extended to other disasters of more immediate concern \citep{teng2021climate}. Hans Jonas stands out for his explicit engagement of the importance of human survival. In Jonas’ philosophy he explores the deeper \textit{a priori} underlying of Immanuel Kant’s categorical imperative \citep{kant1989immanuel}. Kant assumes that humans exist, and upon that a categorical imperative may be built. But the prior axiom that \textit{humans must exist} necessarily comes first \citep{jonas1984imperative}. This is Jonas’s imperative of responsibility: that humankind must exist. No matter what we do we should never endanger this foundational value, because without it the spark of morality is extinguished from the universe \citep{jonas1984imperative}. 
    
\paragraph{Sustainable Intergenerational Existence} Sustainable intergenerational existence is another assumption underlying most philosophies. For example, Confucius explicitly praises family life, quoting the \textit{Book of Odes}: “The happy union with wife and children, is like the music of lutes and harps, when siblings all get along, the harmony is entrancing” \citep{legge1971confucian}. Likewise, Aristotle’s interest in politics manifests first with an interest in family life: “In the first place there must be a union of those who cannot exist without each other; namely, of male and female, that the race may continue…” \citep{aristotle1984complete}. Religions, even more than philosophies, have also certainly emphasized family life--whether Judaism, Christianity, Islam, Hinduism, Buddhism, or other religions--all have had expectations of and ceremonies for marriage, birth, and death. Similarly, contemporary environmental ethics has gained an interest in intergenerational issues, such as intergenerational justice, due to the risk in which future generations now are \citep{meyer2017intergenerational}.
    
\paragraph{Society} The foundation of society is a topic for most social and political philosophies, who emphasize the need for morality, selflessness, justice, and social order. Confucianism’s emphasis on social order and harmony displays a clear emphasis on this value, for example, saying: “men (\textit{sic.}) are close to one another by nature” and for leaders: “Guide them [the people] by virtue, keep them in line with the rites, and they will, besides having a sense of shame, reform themselves” \citep{confucius1979analects}. Aristotle similarly says “the state is a creation of nature and that man is by nature a political animal” thus emphasizing that humans naturally form communities, societies, and polities \citep{aristotle1984complete}. Contemporary anthropology and sociology likewise are very interested in these topics, for example, Donald Brown’s “universal people” use language (presupposing other people to talk to), care about kinship (presupposing the existence of both family and non-family social relationships), are not solitary (spending most if not all of their lives in groups), learn from others (presupposing others to learn from), have division of labor (some people specialize in certain tasks), have morality (presuming others exist to treat well or badly), etc. \citep{brown1991human}.
    
\paragraph{Education} Education tends to be very strongly emphasized by philosophers. Confucius, for example, \textit{Analects} 8:13 says “Have the firm faith to devote yourself to learning” \citep{confucius1979analects}. That Confucius was a teacher demonstrates as well his devotion to education. Plato and Aristotle both emphasized education in their philosophies in their own ways e.g. Plato by \textit{anamnesis} \citep{plato1949meno, plato1972plato}, Aristotle learning-by-doing \citep{aristotle1984complete}, but perhaps even more so in their practices. Each philosopher started his own school for educating pupils, the Academy and the Lyceum, respectively. And, of course, the practical usefulness of school systems are now near-universally agreed upon.
    
\paragraph{Truth} Seeking the truth is a fundamental value of philosophy, the word literally meaning “lover of wisdom.” Aristotle states as the first line in his \textit{Metaphysics} “All men (sic.) by nature desire to know” \citep{aristotle1984complete}. Similarly, the \textit{Analects} of Confucius opens with “Is it not a pleasure, having learned something, to try it out at due intervals?” and Confucius said of himself “I was not born with knowledge but, being fond of antiquity, I am quick to seek it” \citep{lau2000confucius}. It is worth pointing out that contemporary science aims to collect true facts about reality, and the philosophical underpinnings of science provide a theoretical basis, while economic and political desires form a practical basis, for this endeavor.

All of the above five foundations can also be found in one passage of the medieval philosopher Thomas Aquinas in his very brief discussion of the foundations of naturalistic ethics. In this section he lays out a system built upon Aristotle’ notion of there being three layers to the human body and mind: a vegetative soul, the type of which we share with all living things; a sensing soul, the type of which we share with all animals; and a rational soul, the type of which we share only with other rational beings \citep{hicks2015aristotle}. Aquinas updates Aristotle to explain more carefully what would be entailed for sustainable survival of the human species \citep{aquinas2014summa}. While we would now consider both Aristotle and Aquinas to be out-of-date on many issues, at least on this one topic--the core aspects of human nature--Aquinas seems to have struck something significant.

\section{Philosophical Support from Law and Business Practice}

In addition to presenting support for the five values from philosophers, we show that these are widely agreed upon values in our contemporary world as well. While international laws, ideals, and treaties are not typically thought of as philosophy, they certainly express deeper philosophical ideas on values for humanity as a group, such as the importance of cooperation, human equality and dignity, ideas about human rights, and so on \citep{besson2010philosophy}. In this way, we show that the moral foundations presented here align with the “international philosophy” of the United Nations Sustainable Development Goals (SDGs). The SDGs try to protect each of these values in their own ways, specifying practical goals, while assuming the foundations \citep{UnitedNations}. The following classifies the SDGs by the foundational value that it aligns to:

\textbf{Human survival}: 1. No poverty, 2. Zero hunger, 3. Good health and well-being, 6. Clean water and sanitation. 

\textbf{Sustainable intergenerational existence}: 7. Affordable and clean energy, 12. Responsible consumption and production, 13. Climate action, 14. Life below water, 15. Life on land.

\textbf{Society}: 5. Gender equality, 8. Decent work and economic growth, 9. Industry, innovation, and infrastructure, 10. Reduce inequalities, 11. Sustainable cities and communities, 16. Peace, justice, and strong institutions, 17. Partnerships for the goals.

\textbf{Education}: 4. Quality education.

\textbf{Truth}: none explicitly call for this, but \textit{all}, in their desire for metrics and progress, demand it.

Having more goals in one moral value foundation does not imply more importance, only more specification. Just because something is more carefully described does not necessarily imply greater importance, just that more details might be necessary in order to achieve that goal.

Like the UN’s SDGs, international law, human rights discourse, and corporate efforts at technology ethics also embed philosophical values, for example, with Google, IBM, Microsoft, OpenAI, and Salesforce’s principles \citep{GoogleAI, IBM, Microsoft, OpenAICharter, Salesforce}. This shows that corporations are implicitly supporting these five foundational moral values. See table in Appendix \ref{appendix:corp} for details.

\section{Takeaways for Language Model Alignment}

\begin{figure}[t]
    \centering
    \includegraphics[width=0.6\textwidth]{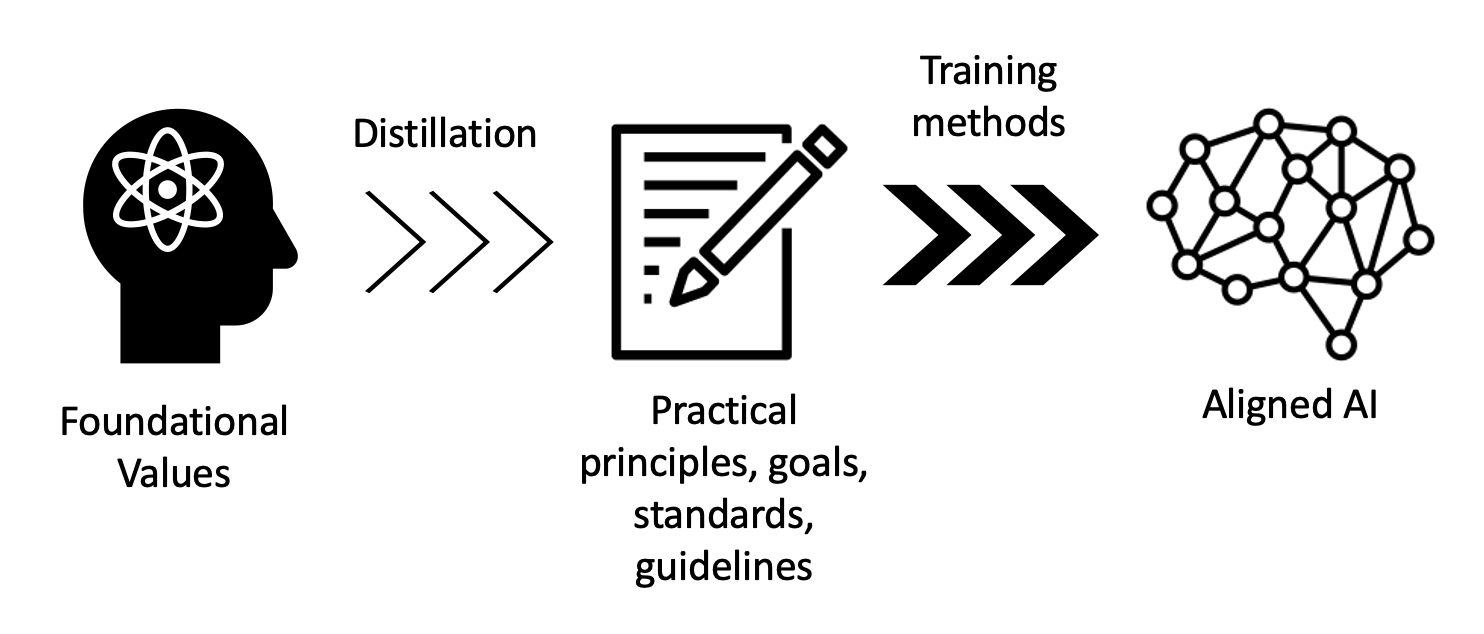}
    \caption{The process of aligning AI systems involves identifying foundational values to distill into mid-level, practical principles, or goals, standards, and guidelines, that can then be used in the training methods of the models.}
    \label{fig:alignment}
\end{figure}

Again, aligning AI agents requires having both the technical methods to encode values into models as well as the values themselves. When looking at large language models, reinforcement learning with human feedback \citep{bai2022training, glaese2022improving}, preference-model pretraining \citep{askell2021general}, and Constitutional AI \citep{bai2022constitutional} are some methods of doing the first part by incorporating human judgements into the training and development process. The foundational values we propose here help provide the second part, although importantly, these technical methods rely on goals, standards, and guidelines that can be used in practice to steer the model. The foundational values are the most abstract level of guidance and therefore require further specification, as it may still be hard to evaluate an LLM output against them, or to use them as guidance when giving feedback for training. As such, we must still bridge the two parts by specifying the foundational values into mid-level principles and action-guiding principles specific to very concrete situations. These specifications are often too detailed to enumerate outside of the context of their use, and so organizations should develop careful procedures for generating these action principles \citep{flahauxethics}. 
\Cref{fig:alignment} shows this process.

By way of example, the values of helpfulness, harmlessness, and honesty \citep{askell2021general} are a step in the process of distilling the foundational values into practical values with which to train models: Honest can be translated to a rule of “Do not lie”, harmless to “Do not do things that harm”, and helpful to “Do things that help”. However, determining whether something is helpful, harmless, or honest depends on how the terms are defined and scoped. By making these values explicit (even if still abstract), models can be made helpful in promoting survival, sustainable intergenerational existence, society, education, and truth; and harmless in terms of not threatening survival, sustainable intergenerational existence, society, education, and truth. Hence, the foundational values provide the means to clarify and specify these concepts. Similarly, any set of practical values selected to build an AI system around should be grounded in foundational values.

\section{Agenda for AI}

Alignment is not only a matter of how the models behave on their own or in a local setting, but also how they will interface with society and the role that they will play. It would not be sufficient to have models built to be aligned to the developer’s intent but still able to be used in harmful ways by other individuals or in other applications. As such, in addition to providing the basis for models themselves, the foundational values act as a \textit{framework} to suggest key areas of consideration for AI practices, the environment in which these models exist, and how the models interface with individuals and society. While many of the following are areas of interest already, the foundational values provide a way to understand how they fit together and might be prioritized as we attempt to address each of them.  

\paragraph{Survival} For survival, we should consider how AI could accelerate catastrophic risks: climate change, nuclear warfare, biotechnology and synthetic biology, nanotechnology, lethal autonomous weapons, etc.. We should also prevent ways that malactors could intentionally use models to harm others, such as through disinformation or infohazards \citep{vallor2018ethics}. Lastly, we should identify and address ways that people could unknowingly harm themselves or others through methods such as misinformation, or enabling followers of apocalyptic cults or terrorist groups \citep{torres2018would}; as technology becomes easier to use and more accessible, these situations become increasingly plausible.

\paragraph{Sustainable Intergenerational Existence} In protecting the next generation, we should not only protect the environment, but also ensure that core human relationships are not being threatened: communities, parent-child relations, partnerships, and family. Although it is again not necessary for every individual to do and these relationships also come in many shapes and forms, the concept and general structure of these relationships play a key role in the preservation of society, as they promote stability, security, and nurturance. Threats in this domain may come from different forms of obsessive use of technology that lead individuals to remove themselves or become distanced from society \citep{fasoli2021overuse}, or ways that AI could encourage destructive behavior towards one’s relationships. 

\paragraph{Society} When considering society, we want to consider how AI is disrupting society. This means looking at changes that AI is bringing on a societal level such as replacing jobs or changing  desired skills for certain jobs \citep{george2023chatgpt}, as well as ways that AI could enable disorder either on the individual or communal level, for example, by enabling or promoting rudeness and rule-breaking \citep{Lee_2023}, or reflecting these traits in outputs. Another dimension of AI’s effect on society is how AI will exist in society. To this end, it is important to consider what policies, regulations, and/or auditing processes \citep{mokander2023auditing} should be in place regarding the usage, deployment, and downstream effects of AI.

\paragraph{Education} Regarding education, we consider both formal education and the broader education of cultural and social norms. With formal education, if students are to become more and more reliant on AI, we should consider how education should adapt to the advent of new ways of learning through AI, whether this means adjusting the content taught or ways of teaching. AI also will have a potentially great impact on the cultivation of moral, societal, and cultural skills, and so warrants close attention and possibly intervention \citep{vallor2015moral}. 

\paragraph{Truth} Lastly, AI should not hinder progress in truth-seeking, either on an individual or societal level. This suggests that, first, models themselves must be truthful–outputs must conform to facts. As we are increasingly relying on and interacting with models, we want to be sure that models do in fact complement our own knowledge and abilities rather than mislead us. This has particularly been a concern with large language models exhibiting hallucinations, producing content that diverges from the user input or contradicts established world knowledge \citep{zhang2023siren}. It should also be emphasized that models do not facilitate the generation or proliferation of mis- or disinformation, in the form of text, images, audio, video, etc \citep{kreps2022all, shu2020combating, bontridder2021role}.

\section{Opportunities for AI}

The above lists the first order of work that must be done such that the five foundational moral values are not threatened by AI; however, there also exist many opportunities to use AI for good in each of these areas, which ultimately have been the appeal of new systems, technologies, and AI abilities–to be a tool to enhance humanity and the society we live in.

\paragraph{Survival} When we consider survival, AI systems can be used to prevent existential and life-threatening risks, such as through automatically detecting extremist users and content \citep{fernandez2021artificial}, and various forms of law enforcement \citep{raaijmakers2019artificial}. There are many opportunities to use AI in the face of natural disasters including floods \citep{nevo2019ml} and earthquakes \citep{devries2018deep}, as well as to assist developing countries in tackling economic and social challenges \citep{kshetri2020artificial}. In healthcare, we have seen AI’s ability to perform medical tasks such as radiology with high accuracy and efficiency \citep{zbontar2018fastmri, jiang2023health}.

\paragraph{Sustainable Intergenerational Existence} AI can support sustainable intergenerational existence by helping to solve long-term environmental problems, such as climate change and threats to biodiversity. At the more immediate and familial level, AI might help by facilitating ways people can meet one another \citep{wang2023algorithmic}. AI should facilitate humans forming relationships with each other, as opposed to attempting to replace or distract from human relationships with itself. More generally, AI systems can promote healthy lifestyles and/or habits so that we can live longer and more fruitful lives, i.e. health monitoring \citep{sujith2022systematic}, which ultimately also aids the long-term survival of humanity, since people are more likely to care about a future in which they will be alive.

\paragraph{Society} There exist many ways that AI can strengthen society. In this regard we can look at the individual level of how AI can make individuals better members of society, and the communal level of how AI can support the structure of society. On the individual level, AI can be aligned towards the immediate values of the rule-of-law, politeness, cooperation, etc. which emphasize interpersonal social order; and more generally values that we identify as important traits for members of society: kindness, compassion, empathy, respect, integrity, responsibility, and so on. On the communal level, there are opportunities to use AI for efficient resource utilization, division of labor, and supply chain management, among other applications.  

\paragraph{Education} The accessibility of information that new AI systems provide suggests that education has great potential for enhancement by AI, just as advancing technologies have been used for education throughout history \citep{raja2018impact}. Intelligent tutoring systems offer flexibility in presentation of material and greater ability to adapt and respond to students’ needs \citep{beck1996applications}, improving educational equity and quality \citep{pedro2019artificial}. AI also offers ways of preserving culture and history to educate future generations, one example being USC’s Dimensions in Testimony, a digital system which enables people to have conversational interactions with Holocaust survivors and other witnesses to genocide \citep{traum2015new}.

\paragraph{Truth} Lastly, as the appeal of AI largely rests on enhancing intelligence, AI has and will continue to aid us in truth-seeking and discovery. We employ AI to do difficult computational problems, model large problems, analyze complex data, and draw inferences from weak patterns. As such, AI has long been an aid in coming up with new ideas and findings. With new language capabilities in large language models, we can now engage with models more easily in creative processes as well, such as writing \citep{chakrabarty2023creativity}.  

It is important that in these situations of well-intended AI use, we do not inadvertently create new problems from AI itself–unfair and biased systems \citep{mehrabi2021survey}, overreliance and resulting deskilling \citep{green2019artificial}, various unintended consequences \citep{amodei2016concrete}, etc. As such, it is important to consider the full range of impacts of employing AI and to fully assess both the short and long-term effects.

\section{Discussion}

In this work we emphasize the need for foundational values in answering the question of "To what values do we align?" and suggest these five foundational values as a potential solution; below we discuss limitations to this proposal. We have not suggested alternative sets of foundational values here, though this is absolutely necessary and we expect much continued discussion on the topic.

\subsection{Anthropocentrism}
This theory is clearly anthropocentric. It focuses on exactly what humans need in order to continue as a species, and any applicability to other entities, natural or artificial, is an instrumental value to protect human existence. However, this ethic can be easily expanded to include other living things: each species has its own set of species-typical foundational moral values, typically, survive and reproduce. If the species is social, then that is added too, and if the species has a culture requiring education (e.g., some kinds of animal language or tool use), then that would be added as well. The next problem is aligning (or in cases of conflict, prioritizing) the competing values of each species \citep{green2020convergences}.

As for the question of whether a conscious AI might supersede humans as a source of moral value this can be considered but should not be considered in a way that undercuts human moral value. As a species-ethic, these are the values for us human beings. An AI agent might have different values, thus risking conflict and presenting danger to both species of moral entity. This is why value alignment is so important in the first place, so that such clashes will be avoided. 

\subsection{How many foundational moral values?}
Perhaps instead of: survive (right now), sustain intergenerational existence (survive long term), live in society (right now), educate young (live in society long term), and seek the truth, there is: survive long term, live in society long term, and seek the truth. Reducing the five values to three is possible, but there is a qualitative difference between surviving right now and surviving for 1000 years, and likewise between living in society right now and living in society for 1000 years. 

If there could be three values instead of five, then why not one or seven? For example, could “seek the truth” or “survive” actually sum up all of the values into one (as Jonas would do with his imperative of responsibility \citep{jonas1984imperative}), since each value necessarily implies the others? Likewise, if one were to derive an ethic from evolutionary theory it might be something like “do not go extinct” which could be a one-rule ethic like Jonas, but would still obscure the values necessary to make this one value sensible. It is also certainly possible for there to be more than five values, but they would likely nest into one of the five we articulate here, and splitting them out might not necessarily elaborate sufficient distinctions to warrant the separation. Still, there are a very large number of values (as human rights discourse exhibits), and the closer we get to practice (and concrete definite actions) the more they multiply, so debate is possible. Nevertheless, we have shown the five here to be foundational, demonstrable, and worthy of serious consideration.

\subsection{How do fundamental values lead to specific actions?}
Another limitation might be the lack of specification from fundamental moral values to immediately usable principles, as one might act upon in a specific case. This problem regarding the application of general rules to specific cases is an old one, dating back to Aristotle, but it should be seen as a challenge, not an impossibility. Despite the difficulty, humans seem to be able to apply general rules to specific cases all the time, which raises the question of how. The answer is that not all values exist on the same level of specificity. Instead they range from fundamental moral values as listed here, to mid-level/intermediate principles of various sorts, down to progressively more and more specific principles that dictate very precise actions such as “in X situation, do Y.” This hierarchy of values is not always clear, but most ethical principles exist in this space, with the extremely general and extremely specific sometimes ignored. There are resources for thinking about the application of moral principles in the AI, technology, bioethical, business, and other contexts \citep{flahauxethics, awad2022computational}.

\subsection{People will have different meanings of the values}
It is clear that different groups and individuals will have different interpretations of each of the five values, even in ways that may conflict. The goal of the foundational values is not to converge towards a consensus on what their specifications should look like in all situations, or on what human flourishing means for every single person; rather the values are open for different meanings and interpretations across societies. But these fundamental values are ones that human groups must all strive for regardless in the effort to align AI, otherwise they cannot last. In the same way that individuals and groups have different views on humanity, they will have different views on the specifications of the foundational values--and yet, in the end, these values must be consistent enough at the global level to permit sustainable human existence \citep{hou2023multi}. This then raises the question as to whether it is possible to truly have AI that is aligned on the global level, or in any scope that encompasses people with conflicting views. We leave this as an open question.

\section{Conclusion}

In conclusion, we argue that alignment efforts necessitate foundational values to align to, which are rooted in the core principles of our human existence. We present five foundational moral values: survival, sustainable intergenerational existence, society, education, and truth. We show that these values are necessary for human existence, and logical via \textit{reductio ad absurdum}. These values find support in global philosophical traditions, the United Nations Sustainable Development Goals, and in various contemporary efforts at tech and AI ethics. These values also present fruitful paths forward in terms of overcoming challenges and finding opportunities. Lastly, while these values have some limitations, these limitations are not so strong as to vitiate the usefulness of this approach.

\begin{ack}
We thank Hannah Rose Kirk for valuable discussion and feedback on the work. BLH is supported by an NSF Graduate Research Fellowship. 
\end{ack}

\bibliography{main}

\appendix
\section{Philosophical Support for the Five Foundational Moral Values}
\label{appendix:phil}
Philosophical support for the five foundational values can be argued by \textit{reductio ad absurdum}:

\begin{enumerate}
    \item Do not go extinct, short term. The opposite would be to value and seek going extinct, which would yield the absurd outcome of undercutting the existence of value itself, since humans are the only known organism capable of moral evaluation. Valuing our own extinction would mean valuing non-valuing, a contradiction.
    \item Do not go extinct, long term. The opposite would be to value and seek going extinct long term, destroying the environment, or not raising a next generation, again leading to extinction and the absurdity of valuing non-valuing. 
    \item Do not live in complete isolation, short term. The opposite would be to value and seek living in isolation, which would again yield the absurd outcome of undercutting the existence of value itself, since humans cannot live in complete isolation even short-term. The current generation would die from breakdown of division of labor and supply chains, again resulting in extinction and valuing non-valuing.
    \item Do not live in complete isolation, long term. The opposite would be to value and seek living in isolation, long term, which would again yield the absurd outcome of undercutting the existence of value itself, since humans cannot live without care or education. The current generation would die and we would not have another, to value this would again lead to valuing non-valuing.
    \item Do not seek falsehood. The opposite would be to value and seek falsehood, which would yet again yield the absurd outcome of undercutting the existence of value itself, since seeking falsehoods would entail seeking both theoretical and practical falsehoods, and practical falsehoods would result in death in relatively short order, as people eat non-foods, drink poisons, die of exposure, jump off cliffs, etc.
\end{enumerate}
 
\section{Corporate Values and the Five Foundational Moral Values}
\label{appendix:corp}
Corporate efforts at technology ethics also embed philosophical values. \Cref{tab:corp_values} shows examples, categorizing each corporation’s principles into the foundational value that they align with. 

We observe here that all the values except education are represented, though only OpenAI and Salesforce hit sustainable/long-term human existence. Society is the most highly represented value, as might be expected for corporations seeking social benefit. Also, as might be expected for corporations, short term is emphasized over long term, which might represent an opportunity for improvement. The lack of representation of education as a value probably reflects the assumption by these companies that educated workers will always be available, whether they explicitly value them or not, or that these companies' educational efforts are separate from the efforts that wrote their AI principles.

In any case, this helps to show that corporations are implicitly supporting these five foundational moral values. But even then, opportunities for improvement are apparent, and differences in emphasis, likely partially reflecting corporate cultures. Additionally, while these values might be explicitly stated, companies may not have found a way to transfer these values into the models and systems they're building, hence the need for both examining the foundational level under these corporation’s intermediate principles, and more specific action principles beyond mere theory, where “the rubber hits the road” \citep{flahauxethics}.

\definecolor{Gray}{gray}{0.8}
\definecolor{LightCyan}{gray}{0.9}

\begin{center}

\begin{table}
    \centering
    \begin{tabular}{|p{1.6cm}|p{1.8cm}|p{1.8cm}|p{2.5cm}|p{1.8cm}|p{1.8cm}|}
        \hline
        \rowcolor{Gray}
         &  \textbf{Survival} & \textbf{Sustainable Intergenerational Existence} & \textbf{Society} & \textbf{Education} & \textbf{Truth} \\
         \hline
         \cellcolor{LightCyan} OpenAI: Charter Principles \citep{OpenAICharter} &  & Long-term survival & Broadly distributed benefits, Cooperative orientation &  & Technical leadership \\
         \hline
         \cellcolor{LightCyan} Google: AI Principles \citep{GoogleAI} & Be built and tested for safety &  & Be socially beneficial, Avoid creating or reinforcing unfair bias, Be accountable to people, Incorporate privacy design principles, Be made available for uses that accord with these principles &  & Upload high standards of scientific excellence \\
         \hline
         \cellcolor{LightCyan} Microsoft: Responsible AI Principles \citep{Microsoft} & Reliability \& Safety &  & Fairness, Privacy \& Security, Inclusiveness, Accountability &  & Transparency \\
         \hline
         \cellcolor{LightCyan} IBM: AI Pillars \citep{IBM} & Robustness &  & Fairness, Privacy &  & Explainability, Transparency \\
         \hline
         \cellcolor{LightCyan} Salesforce: Core Values \& Ethical Use Guiding Principles \citep{Salesforce} & Safety & Sustainability & Customer success, Equality, Human rights, Privacy, Inclusion &  & Trust, Innovation, Honesty \\
         \hline
    \end{tabular}
    \caption{Classification of principles from selected corporations into the five foundational values}
    \label{tab:corp_values}
\vspace{-5pt}
\end{table}
\end{center}


\end{document}